\documentclass[reqno]{amsart}
\usepackage{amsmath}
\usepackage{amsfonts}
\usepackage[dvips]{graphics}
\usepackage{epsfig}
\newtheorem{theorem}{Theorem}[section]
\usepackage[T1]{fontenc}
\usepackage[utf8]{inputenc} 

\begin{document}
\title{The infinite volume limit of Ford's alpha model}
\author{Sigurður Örn Stefánsson
\address{Science Institute, University of Iceland, Dunhaga 3, 107 Reykjavík, Iceland}}
\maketitle
\begin {center}
\today 
\end {center}
\begin{abstract}
We prove the existence of a limit of the finite volume probability measures generated by tree growth rules in Ford's alpha model of phylogenetic trees. The limiting measure is shown to be concentrated on the set of trees consisting of exactly one infinite spine with finite, identically and independently distributed outgrowths. 
\end{abstract}

\section{Introduction}

Graphs are used in many fields of science to describe relationships between individuals and to model actual physical objects. The former case includes social networks~\cite{albert0}, phylogenetic trees~\cite{Aldous,Ford,fragmentation}, the world-wide web~\cite{albert1} and much more. The latter case includes discrete objects such as macromolecules~\cite{RNAfolding} and branched polymers~\cite{albert0}. The graphs can also serve as discrete approximations to inherently continuous objects, an example of this being triangulation of manifolds in quantum gravity, see e.g.~\cite{book}. 

Random graphs are commonly used to describe real deterministic networks.  Interactions and relations in the networks can be complicated but their characteristics are in some cases captured by random graph models, defined by simple rules which are motivated by the nature of the real network.
The alpha model, introduced by D. Ford in~\cite{Ford}, is an example of a random graph model, intended to describe phylogenetic trees. It is a one parameter model of randomly growing, rooted, planar, binary trees with the following growth rules. Start from a single rooted edge and from a tree on $n$ leaves, select individual internal edges with probability weight $\alpha$ and individual leaves with probability weight $1-\alpha$ where $0\leq\alpha\leq 1$. Graft a new leaf to a selected edge and thus generate a tree on $n+1$ leaves, see Fig.~\ref{F:grafting}. 

Ford proved that the model is Markovian self-similar which means informally that a subtree below an edge is distributed identically to the whole tree, a more precise definition will be given in the main section. He also showed that typical distances in the trees scale as $n^{\alpha}$ with the system size $n$. The Hausdorff dimension of a randomly growing tree is defined to be $d_H$ given that typical distances scale as $n^{1/d_H}$. Thus, in the alpha model $d_H = 1/\alpha$. 

In a recent paper~\cite{fragmentation} the continuum limit of the model has been established in the context of fragmentation processes~\cite{fbook}. A generalization to multinary trees is introduced in~\cite{alphagamma} in the alpha-gamma model where in addition to the growth rules of the alpha model, edges can be grafted onto vertices, increasing their degree. The alpha-gamma trees are shown to be Markovian self-similar and a continuum limit is established. 

Our motivation to study the alpha model comes from the fact that it is a certain limiting case of a model of random trees which grow by vertex splitting, introduced in~\cite{vs} where the relation is explained. In general the vertex splitting model does not share some of the technically convenient properties of the alpha model such as Markovian self-similarity, and it is more difficult to do exact calculations. The hope is that some of these properties might hold asymptotically for large trees and therefore a good understanding of the alpha model could be helpful. 

The purpose of this paper is to establish convergence of the finite volume measures generated by the alpha model to a measure on infinite trees. For $0 < \alpha \leq 1$, the infinite measure is shown to be concentrated on the set of trees consisting of exactly one infinite path from the root to infinity (spine) with finite, identically and independently distributed outgrowths. 

\section {Convergence of the finite volume measures}

We start with a few definitions before presenting the model. In this paper we only consider rooted, binary, planar trees. Rooted means that we mark a single vertex of degree 1, binary means that vertices are only allowed to have degree 1 or 3 and the planarity condition distinguishes between left and right branchings. The root and vertices of degree 3 will be referred to as internal vertices and vertices of degree 1, besides the root, will be referred to as leaves.  Denote the set of trees on $n$ leaves by $T_n$ and denote the set of all finite or infinite trees by $T$. 

The alpha model is defined by probability distributions $\pi_{\alpha,n}$ on $T_n$, for $n \geq 1$, constructed in the following recursive way. Assign probability one to the unique trees in $T_1$ and $T_2$. Given $\pi_{\alpha,n}$ for some $n\geq 2$, $\pi_{\alpha,n+1}$ is generated by first selecting a tree $\tau \in T_n$ according to $\pi_{\alpha,n}$. Next an individual edge $(a,b)$ is selected from $\tau$ with probability $\alpha/(n-\alpha)$ if $a$ and $b$ are internal vertices and with probability $(1-\alpha)/(n-\alpha)$ if one is an internal vertex and the other a leaf. The edge $(a,b)$ is removed from $\tau$ and two new vertices $c$ and $d$ are introduced along with the edges $(a,c)$, $(c,b)$ and $(c,d)$. 
\begin{figure} [!h]
\centerline{\scalebox{1}{\includegraphics{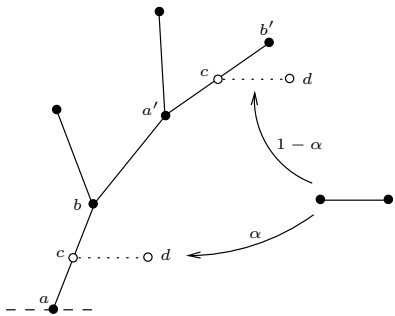}}}
\caption{The grafting process. The link $(a,b)$ is selected with probability weight $\alpha$ and the link $(a',b')$ is selected with probability weight $1-\alpha$. The selected link is removed, two new vertices $c$ and $d$ and three new links are added as shown in the figure. In this example, $a$ is the root which is indicated by a dashed line.} \label{F:grafting}
\end{figure}
Equal probability is assigned to left and right branching of the new edge $(c,d)$. One can think about this procedure as grafting a new edge to an existing edge in $\tau$, see Fig.~\ref{F:grafting}. 
The probability of a tree $\tau' \in T_{n+1}$ is thus given by
\begin {equation}
 \pi_{\alpha,n+1}(\tau') = \sum_{\tau\in T_n} \pi_{\alpha,n}(\tau) \mathbb{P}(\tau\rightarrow\tau')
\end {equation}
where $\mathbb{P}(\tau\rightarrow\tau')$ is the probability of growing the tree $\tau'$ from $\tau$ by the grafting process.

\begin{figure} [!h]
\centerline{\scalebox{1}{\includegraphics{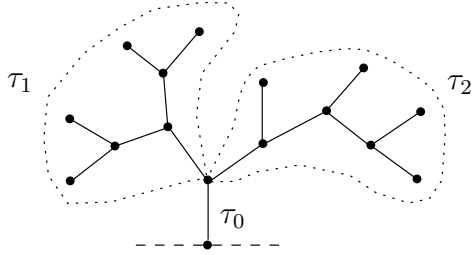}}}
\caption{An example of a tree $\tau_0$ which has a root indicated by the dashed line. The tree $\tau_0$ branches at the nearest neighbour of the root to two subtrees, $\tau_1$ to the left and $\tau_2$ to the right as is indicated by the dotted lines.} \label{branching}
\end{figure}

The model has a property called Markovian self-similarity~\cite{Ford} which is essential in the inductive proof of the theorem in this paper.  Markovian self-similarity means that there exists a function $q_\alpha(\cdot,\cdot)$ such that for every finite tree $\tau_0$ which branches at the nearest neighbour of the root to a left tree $\tau_1$ and a right tree $\tau_2$ (see Fig. \ref{branching}) the following holds

\begin{equation}
 \pi_{\alpha,|\tau_0|}(\tau_0) = q_\alpha(|\tau_1|,|\tau_2|)\pi_{\alpha,|\tau_1|}(\tau_1) \pi_{\alpha,|\tau_2|}(\tau_2)
\end{equation}
where $|\tau|$ denotes the number of leaves in a tree $\tau$. In words, this says that  $q_\alpha(n_1,n_2)$ is the probability of a tree branching to subtrees of sizes $n_1$ and $n_2$. Furthermore, given that the subtrees are of these sizes they are distributed independently by $\pi_{\alpha,n_1}$ and $\pi_{\alpha,n_2}$.
The function $q_\alpha$ is explicitly known~\cite{Ford} and is given by
\begin {eqnarray*}
 q_\alpha(n_1,n_2) &=&  \frac{n!\Gamma_\alpha(n_1)\Gamma_\alpha(n_2)}{n_1! n_2!	\Gamma_\alpha(n)}\left(\frac{\alpha}{2 }+\frac{(1-2\alpha)n_1n_2}{n(n-1)}\right)
\end {eqnarray*}
where $n = n_1+n_2$,	
\begin {equation}
\Gamma_{\alpha}(n) = (n-1-\alpha)(n-2-\alpha)\cdots(2-\alpha)(1-\alpha), \quad \text{and} \quad \Gamma_{\alpha}(1) = 1.
\end {equation}

Before proceeding to the theorem we give a short explanation of what is meant by convergence of probability measures. For a tree $\tau \in T$ let $B_R(\tau)$ be the subtree of $\tau$ which is spanned by the vertices at distance less than or equal to $R$ from the root of $\tau$. Define a metric $d$ on $T$ by
\begin {equation}
d(\tau,\tau') = \inf\left\{\frac{1}{1+R}~\Big|~B_R(\tau) = B_R(\tau')\right\}.
\end {equation}
For some properties of the metric space $(T,d)$ see~\cite{billingsley,bergfinnur}. We will establish weak convergence, as $n\rightarrow\infty$ of the measures $\pi_{\alpha,n}$ viewed as probability measures on $T$, to a probability measure $\pi_{\alpha}$. This means that for all bounded functions $f$ which are continuous in the topology generated by the metric $d$
\begin {equation}
\int_T f(\tau)d\pi_{\alpha,n} \longrightarrow \int_T f(\tau)d\pi_{\alpha}, \quad\quad \text{as $n \longrightarrow \infty$.}
\end {equation}
 
\begin{figure} [!h]
\centerline{\scalebox{1}{\includegraphics{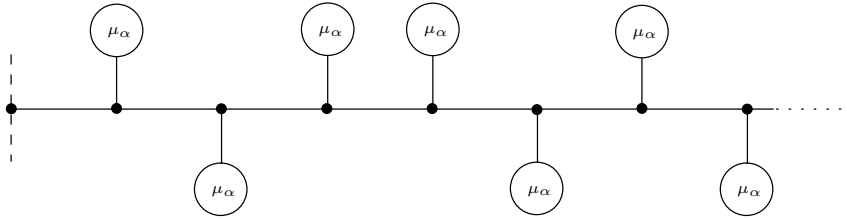}}}
\caption{The infinite spine with finite $\mu_\alpha$--outgrowths.} \label{f:spine}
\end{figure}

\begin {theorem}
Let $0 < \alpha \leq 1$. The measures $\pi_{\alpha,n}$, viewed as probability measures on $T$, converge weakly, as $n \longrightarrow \infty$, to a probability measure $\pi_\alpha$ on infinite trees which is concentrated on the set of trees with one infinite rooted spine with finite outgrowths i.i.d. by 
\begin {equation}
 \mu_\alpha(\tau) =  \frac{\alpha\Gamma_\alpha(|\tau|)}{|\tau|!}\pi_{\alpha,|\tau|}(\tau).
\end {equation}
The probabilities of right and left branching of outgrowths are equal (see Fig.~\ref{f:spine}).
\end {theorem}
{\it Proof.} We call the maximum graph distance from the root to a leaf in a tree, the height of the tree. Let $T^{(R)}$ be the set of rooted trees of height $R$. The metric space $(T,d)$ is compact and therefore it is sufficient to show that for any $R \geq 1$ and any $\tau_0 \in T^{(R)}$ the sequence
\begin {equation}
 \pi_{\alpha,n}(\{\tau | B_R(\tau) = \tau_0\}) =: \pi_{\alpha,n}^{(R)}(\tau_0)
\end {equation}
 converges to a limit $\pi_\alpha^{(R)}(\tau_0)$ as $n\longrightarrow\infty$~\cite{bergfinnur}. We show this by induction on $R$. For $R = 1$ this is trivial since  $B_1(\tau) \in T^{(1)}$ for all $\tau$. Now assume that for some $R$ and all $\tau \in T^{(R)}$,  $\pi_{\alpha,n}^{(R)}(\tau)$ converges as $n\longrightarrow \infty$. Choose a tree $\tau_0 \in T^{(R+1)}$ and without loss of generality, assume it branches at the nearest neighbour of the root to a left tree $\tau_1 \in T^{(R)}$ and a right tree $\tau_2\in T^{(S)}$ (see Fig.~\ref{branching}) where $S \leq R$. Then

\begin {eqnarray} \nonumber
 \pi^{(R+1)}_{\alpha,n}(\tau_0) &=& \sum_{n_1+n_2=n}q_\alpha(n_1,n_2)\pi^{(R)}_{\alpha,n_1}(\tau_1) \pi^{(R)}_{\alpha,n_2}(\tau_2) \\ \nonumber
&=& \frac{n!}{\Gamma_\alpha(n)}\Big(\frac{\alpha}{2}\sum_{n_1+n_2=n}\frac{\Gamma_\alpha(n_1)\Gamma_\alpha(n_2)}{n_1!n_2!}\pi^{(R)}_{\alpha,n_1}(\tau_1) \pi^{(R)}_{\alpha,n_2}(\tau_2) \\  \nonumber
&& +~\frac{1-2\alpha}{n(n-1)}\sum_{n_1+n_2=n}\frac{\Gamma_\alpha(n_1)\Gamma_\alpha(n_2)}{(n_1-1)!(n_2-1)!}\pi^{(R)}_{\alpha,n_1}(\tau_1) \pi^{(R)}_{\alpha,n_2}(\tau_2)\Big). \\ \label{PRec}
\end {eqnarray}
If $S < R$ then $\pi^{(R)}_{\alpha,n_2}(\tau_2)=0$ when $n_2 > \ell(\tau_2)$ and it is obvious from the induction hypothesis that $\pi^{(R+1)}_{\alpha,n}(\tau_0)$ converges. Therefore assume that $S = R$. 

Note that in (\ref{PRec}) it always holds that either $n_1 \leq n-1$ and $n_2 \leq n$ or $n_2 \leq n-1$ and $n_1 \leq n$. Therefore we have the upper bound

\begin {eqnarray*}
\pi^{(R+1)}_{\alpha,n}(\tau_0) \leq \frac{n!}{\Gamma_\alpha(n)}\sum_{n_1+n_2=n}\frac{\Gamma_\alpha(n_1)\Gamma_\alpha(n_2)}{n_1!n_2!}.
\end {eqnarray*}
Terms in the sums in (\ref{PRec}) for which $n_1 \geq \frac{n}{2}$ and $n_2 > A$ or $n_2 \geq \frac{n}{2}$ and $n_1>A$ where $A>1$ is some constant are therefore bounded from above by

\begin {eqnarray} \nonumber
\frac{2n!}{\Gamma_\alpha(n)}\sum_{\substack{n_1+n_2=n\\ n_1\geq n/2, n_2 > A}}\frac{\Gamma_\alpha(n_1)\Gamma_\alpha(n_2)}{n_1!n_2!} \leq \frac{2n!\Gamma_\alpha([n/2])}{\Gamma_\alpha(n)[n/2]!}\sum_{n_2=A}^\infty\frac{\Gamma_\alpha(n_2)}{n_2!} \\ \label{Atozero}
\leq C \sum_{n_2=A}^\infty\frac{\Gamma_\alpha(n_2)}{n_2!} \quad \substack{\longrightarrow \\  A \longrightarrow \infty} \quad 0
\end {eqnarray}
where $C$ is a constant. The remaining contribution to (\ref{PRec}) is from terms where $n_1 \geq \frac{n}{2}$ and $n_2 < A$ or $n_2 \geq \frac{n}{2}$ and $n_1 < A$. Notice that the second term in that contribution to (\ref{PRec}) will be of one order lower in $n$ than the first term. Therefore it is enough to show that the first term converges as $n \rightarrow \infty$ since then the second term clearly converges to zero. The contribution to the first term is
\begin {eqnarray} \nonumber
 \frac{n!}{\Gamma_\alpha(n)}\frac{\alpha}{2}\sum_{i=1}^{2}\sum_{\substack{n_1+n_2=n\\n_j \leq A, j\neq i}}\frac{\Gamma_\alpha(n_1)\Gamma_\alpha(n_2)}{n_1!n_2!}\pi^{(R)}_{\alpha,n_1}(\tau_1) \pi^{(R)}_{\alpha,n_2}(\tau_2) \\ \nonumber
\quad \substack{\longrightarrow \\ n \longrightarrow \infty} \quad  \frac{1}{2} \sum_{\substack{i=1\\j \neq i}}^{2}\pi_\alpha^{(R)}(\tau_i)\sum_{m=1}^{A} \frac{\alpha \Gamma_\alpha(m)}{m!}\pi_{\alpha,m}^{(R)}(\tau_j) \\ \label{formoflimit}
\quad \substack{\longrightarrow \\ A \longrightarrow \infty} \quad  \frac{1}{2} \sum_{\substack{i=1\\j \neq i}}^{2}\pi_\alpha^{(R)}(\tau_i)\sum_{m=1}^{\infty} \frac{\alpha \Gamma_\alpha(m)}{m!}\pi_{\alpha,m}^{(R)}(\tau_j).
\end {eqnarray}
In the first step we used the induction hypothesis. This is the limit of $\pi^{(R+1)}_{\alpha,n}(\tau_0)$ as $n \longrightarrow \infty$. The fact that the sum in (\ref{Atozero}) converges to zero as $A\rightarrow \infty$ proves that the measure is concentrated on the set of trees with exactly one infinite spine. The last sum in (\ref{formoflimit}) shows that the distribution of the finite outgrowths is given by $\mu_\alpha$. \begin{flushright} $\square$
\end{flushright}

\section{Conclusions}
We have shown that the finite volume measures $\pi_{\alpha,n}$ generated by the growth rules of Ford's alpha model converge, as $n \rightarrow \infty$, to a measure on infinite trees. The limiting measure is concentrated on the set of trees consisting of exactly one infinite spine with finite outgrowths, independently distributed by $\mu_\alpha$. The emergence of a single spine is well known from models of size conditioned critical Galton Watson trees~\cite{sdgt}. The case $\alpha = 1/2$ is in fact a special case of such a tree.  However, in the vertex splitting model it is possible that an infinite number of spines emerge. This happens for example in the special case of the preferential attachment model~\cite{massdist} and in the case $\alpha=0$ in the alpha model. In both these cases the Hausdorff dimension is infinite. One interesting question is whether a finite Hausdorff dimension is equivalent to the emergence of a single spine and whether an infinite Hausdorff dimension is equivalent to the existence of infinite number of spines in the vertex splitting model.

An obvious next step is to use the formula for the limiting measure to calculate some global properties of the alpha trees such as the Hausdorff dimension and the spectral dimension. The Hausdorff dimension of an infinite random tree given by a probability distribution $\nu$ is defined as $d_H$ if
\begin {equation}
 \langle V_R \rangle_\nu \sim R^{d_H}
\end {equation}
where $V_R(\tau)$ is the number of edges in a ball $B_R(\tau)$ and $\langle \cdot \rangle_\nu$ denotes expectation with respect to $\nu$. The above definition should coincide with the one given by the scaling of a typical distance in a finite tree as discussed in the introduction. This will be checked explicitly in a forthcoming paper.

The spectral dimension of an infinite random tree as above is defined as $d_s$ if 
\begin {equation}
 \langle p(t) \rangle_\nu \sim t^{-d_s/2}
\end {equation}
where $p_\tau(t)$ is the probability that a simple random walk which starts at the root of a tree $\tau$ at time $t=0$ is back at the root at time $t$. The techniques used in~\cite{sdgt} give a way to estimate the spectral dimension of the alpha model from knowledge of the large $R$ behaviour of the quantities $\langle |B_R| \rangle_{\mu_\alpha}$, $\mu_\alpha\{\tau | \text{ height of $\tau > R$}\}$ and $\langle |B_R|^{-1} \rangle_{\pi_\alpha}$. Using the formula for the limiting measure and the Markovian self-similarity properties of the outgrowths one can write recursion equations for generating functions of these quantities. Preliminary results indicate that indeed $d_H = 1/\alpha$ in agreement with the finite scaling definition and $d_s = 2/(1+\alpha)$. In the case $\alpha = 1$ this is trivially true and in the case $\alpha = 1/2$ the result is known to be true by connection to Galton Watson trees~\cite{sdgt}. For other values of $\alpha$ this has not yet been proven.
\\
\medskip

\noindent {\bf Acknowledgment.}  This work is supported by the Eimskip Research Fund at the University of Iceland. I would like to thank Thordur Jonsson, François David and Mark Dukes for helpful discussions and valuable comments.

\end{document}